\documentclass[aps,pra,amsmath,amssymb,showpacs,twocolumn]{revtex4-1}
\usepackage{txfonts}
\usepackage{epsfig,bm,dcolumn}
\usepackage{graphicx}
\usepackage{color}
\usepackage{overpic}

\begin{document}
\title{Thermal Charm and Charmonium Production in Quark Gluon Plasma}
\author{Kai Zhou$^{1,2,3}$, Zhengyu Chen$^1$, Carsten Greiner$^2$ and Pengfei Zhuang$^1$}
\affiliation{$^1$ Physics Department, Tsinghua University and Collaborative Innovation Center of Quantum Matter, Beijing 100084, China\\
             $^2$ Institute for Theoretical Physics, Johann Wolfgang Goethe-University Frankfurt, Max-von-Laue-Strasse 1, 60438 Frankfurt am Main, Germany\\
             $^3$ Frankfurt Institute for Advanced Studies, Ruth-Moufang-Str.1, 60438 Frankfurt am Main, Germany}
\date{\today}
\begin{abstract}
We study the effect of thermal charm production on charmonium regeneration in high energy nuclear collisions. By solving the kinetic equations for charm quark and charmonium distributions in Pb+Pb collisions, we calculate the global and differential nuclear modification factors $R_{AA}(N_{part})$ and $R_{AA}(p_t)$ for $J/\psi$s. Due to the thermal charm production in hot medium, the charmonium production source changes from the initially created charm quarks at SPS, RHIC and LHC to the thermally produced charm quarks at Future Circular Collider (FCC), and the $J/\psi$ suppression ($R_{AA}<1$) observed so far will be replaced by a strong enhancement ($R_{AA}>1$) at FCC at low transverse momentum.
\end{abstract}
\maketitle

Statistical Quantum Chromodynamics (QCD) predicts that, a strongly interacting matter will undergo a deconfinement phase transition
from hadron matter to quark matter at finite temperature and density. It is expected that, this new state of matter, the so-called quark gluon plasma (QGP),
can be created by liberating quarks and gluons from hadrons through high energy nuclear collisions. Since the QGP can only exist in the initial period and cannot be directly observed in the final state of the collisions, one needs sensitive probes to demonstrate the formation of this new state. $J/\psi$ suppression has long been considered as such a probe since the original work of Matsui and Satz~\cite{satz}, and many progresses have been achieved both experimentally and theoretically, see for instance the recent review paper \cite{review,review_zhuang}. While the charmonium production mechanism changes from initial production at SPS energy~\cite{hufner,vogt,zhu} to initial production plus regeneration at RHIC and LHC energies~\cite{pbm,thews,rapp,yan,taesoo,strikland,blaizot,katz}, the charm quarks are all from the initial production.

Recently, the Future Circular Collider (FCC) at CERN is proposed to push the energy frontier beyond LHC, which includes the plan of Pb+Pb collision at $\sqrt{s_{NN}}=39$ TeV~\cite{fcc}. What would we expect about the charmonium production at this new energy regime? Since a much more hot medium will emerge at FCC, gluons and light quarks inside the medium would be more energetic and denser. Therefore, the thermal production of charm quarks via gluon fusion and quark and anti-quark annihilation may have a sizeable effect on charmonium regeneration. For the in-medium charm quark production, there are already many studies, by considering leading order~\cite{lo1,lo2,lo3} and including next to leading order~\cite{zhang} QCD processes. Taking into account the quadratic dependence of the charmonium regeneration on charm quark density, we expect that, the extra increase of charm quark pairs via the thermal production in QGP will obviously enhance the charmonium yield at FCC. Since the very hot medium can eat up almost all the initially produced charmonia, the regeneration becomes the only source of the finally observed soft charmonia. This makes $J/\psi$ more effective to probe the medium properties. In this paper, we focus on the effect of thermal charm production on charmonium production in heavy ion collisions at LHC and FCC energies.

The full information of charm quarks in medium is contained in their distribution function $f_c(t,{\bf x},{\bf p})$ in phase space, its momentum integration is the number density $n_c(t,{\bf x})=\int d^3{\bf p}/(2\pi)^3f_c(t,{\bf x},{\bf p})$. When charm quarks approach kinetic equilibrium with the medium significantly fast in high energy nuclear collisions, only the evolution of the chemical abundance needs to be considered. By integrating out
the charm quark momentum assuming thermal distribution, the Boltzmann equation for $f_c$ becomes the rate equation for $n_c$,
\begin{equation}
\label{rate1}
\partial_\mu n_c^\mu=r_{gain}-r_{loss},
\end{equation}
where $n_c^\mu=n_c(1,{\bf v})$ is the charm current with medium velocity ${\bf v}$, and the gain and loss terms $r_{gain}$ and $r_{loss}$ on the right hand side are respectively the charm quark production and annihilation rates inside QGP. The rates can be calculated through perturbative QCD.

It is convenient to use
the Lorentz covariant variables $\eta=1/2\ln((t+z)/(t-z))$ and $\tau=\sqrt{t^2-z^2}$ to replace time $t$ and longitudinal coordinate $z$. By using $\partial_t=\cosh\eta \partial_\tau-\sinh\eta/\tau \partial_\eta$ and $\partial_z=-\sinh\eta \partial_\tau+\tau\cosh\eta \partial_\eta$, the rate equation is expressed as
\begin{equation}
\label{rate2}
\frac{1}{\cosh\eta}\partial_{\tau}n_c+\nabla_{T}\cdot (n_c{\bf v}_T)+\frac{1}{\tau \cosh\eta}n_c=r_{gain}-r_{loss}
\end{equation}
with transverse medium velocity ${\bf v}_T$.

From the experimentally observed large quench factor~\cite{quench1,quench2} and elliptic flow~\cite{dv1,dv2} for open charm mesons at RHIC and LHC energies, we assume that charm quarks are kinetically equilibrated with the medium during the whole evolution~\cite{taesoo2}. Therefore, the longitudinal motion of charm quarks will be consistent with the medium's Bjorken expansion~\cite{bjorken} in mid rapidity region. To make this explicitly, we set $n_c=\rho_c/\tau$ with $\rho_c(\tau,{\bf x}_T)$
being the charm quark number density in transverse plane and controlled by the reduced rate equation at mid rapidity,
\begin{equation}
\label{rate3}
\partial_{\tau}\rho_c+\nabla_T\cdot (\rho_c{\bf v}_T)=\tau(r_{gain}-r_{loss}).
\end{equation}
Taking into account the nuclear geometry and nuclear shadowing effect on parton distributions in nuclei, the initial condition at time $\tau_0$ for the rate equation in nuclear collisions at fixed impact parameter ${\bf b}$ can be written as,
\begin{eqnarray}
\label{initial}
\rho_c(\tau_0,{\bf x}_T|{\bf b}) &=& \frac{d\sigma_{c\bar c}}{d\eta}T_{A}({\bf x}_T)T_B({\bf x}_T-{\bf b})\nonumber\\
&&\times\mathcal{R}_g(x_1,{\bf x}_T)\mathcal{R}_g(x_2,{\bf x}_T-{\bf b}),
\end{eqnarray}
where $d\sigma_{c\bar c}/d\eta$ is the rapidity distribution
of charm quark production cross section in p+p collisions, and $T_A$ and $T_B$ are the thickness functions at transverse coordinate ${\bf x}_T$ and ${\bf x}_T-{\bf b}$ for the two colliding nuclei. From the FONLL~\cite{fonll} simulation, we extract $d\sigma_{c\bar c}/d\eta|_{\eta=0}=0.7,\ 1.0$ and $2.5$ mb
at colliding energy $\sqrt{s_{NN}}=2.76,\ 5.5$ and $39$ TeV. Considering gluon fusion as the dominant process of charm quark production in high energy p+p collisions, the cross section is multiplied by the
shadowing modification factors $\mathcal{R}_g(x_1,{\bf x}_T)$ for one gluon and $\mathcal{R}_g(x_2,{\bf x}_T-{\bf b})$ for the other to include the nuclear shadowing effect in A+B collisions, where $x_1$ and $x_2$ are the averaged gluon momentum fractions which can be estimated by $x=2\sqrt{m_c^2+\langle p^2_T \rangle}/\sqrt{s_{NN}}$ with averaged charm quark transverse momentum~\cite{fonll}. The space dependence of the shadowing factor is taken as a linearized form~\cite{vogt2},
$\mathcal{R}_g(x,{\bf x}_T)=1+c T_A({\bf x}_T)$. From the normalization condition one can get $c=Z_A(R_g-1)/T_{AB}({\bf 0})$, where $Z_A$ is the nuclear mass number, $T_{AB}({\bf b})=\int d^2{\bf x}_T T_A({\bf x}_T)T_B({\bf x}_T-{\bf b})$ is the nuclear geometry factor, and
$R_g(x)$ is the space independent shadowing factor which is taken from the EKS98 model~\cite{eks98} in the present study .

Now we turn to the loss and gain terms for thermal charm production in medium. The general Lorentz-invariant form for a $2\to 2$ process with initial particles $1$ and $2$ is
\begin{equation}
\label{rate4}
r_{12}=\frac{dn}{d^4x}=\frac{1}{\nu}\int\frac{d^3{\bf p}_1}{(2\pi)^32E_1}\frac{d^3{\bf p}_2}{(2\pi)^32E_2}4F_{12}\sigma_{12}f_1f_2,
\end{equation}
where $\nu$ is the number of identical particles in the initial state, $F_{12}=\sqrt{(p_1\cdot p_2)^2-m_1^2m_2^2}$ is the kinetic flux factor,
$\sigma_{12}$ is the corresponding cross section, and $f_{1,2}$ are the distribution functions of the initial particles $1$ and $2$. For the charm pair production, we take the NLO cross section derived by Mangano, Nason, and Ridolfi (MNR-NLO)~\cite{nason,mnr} and choose the QCD running coupling constant $\alpha_s$ at the renormalization scale $\mu=m_c$.
For initial gluons and light quarks, we take their thermal masses~\cite{mass} $m^2_g = (N_c+N_f/2)g^2T^2/6$ and $m^2_q = (N^2_c-1)/(8N_c)g^2T^2$, where $N_c=3$ is the number of colors and $N_f=3$ the number of flavors. Most of the nonperturbative dynamics is contained in the temperature dependent coupling $g(T)$.
We take it from Ref.\cite{plumari} obtained by fitting the lattice QCD thermodynamics.
Using fully thermal and chemical equilibrium distributions $f_g^{eq}$ and $f_q^{eq}$ for gluons and light quarks, one obtains the in-medium charm pair production rate $r_{gain}=r_{gg\to c\bar c (g)}+r_{q\bar q\to c\bar c (g)}$.

When charm quarks are dense enough, their annihilation starts to reduce the charm quark population.
From the detailed balance between the production and annihilation processes, we can get the annihilation cross section. For the annihilation rate $r_{loss}=r_{c\bar c(g)\to gg}+r_{c\bar c(g)\to q\bar q}$, we need the charm and anti-charm quark distributions $f_c$ and $f_{\bar c}$. Taking into account the strong interactions between charm quarks and the medium, we assume charm quark thermalization and take $f_{c(\bar c)} = n_{c(\bar c)}/n_{c(\bar c)}^{eq}f_{c(\bar c)}^{eq}$, where $f^{eq}_{c(\bar c)}$ are the kinetically thermalized distribution functions for charm and anti-charm quarks and $n_{c(\bar c)}/n^{eq}_{c(\bar c)}=\int d^3{\bf p}/(2\pi)^3f_{c(\bar c)}({\bf p})/\int d^3{\bf p}/(2\pi)^3f_{c(\bar c)}^{eq}({\bf p})$
are just the charm quark fugacity factors $\gamma_{c(\bar c)}$ which measure how far the distributions are from the chemical equilibrium.

For the $3\to 2$ annihilation processes which are at next to leading order, there is one more momentum integration for the initial gluon in the loss rate. We can effectively absorb this into the annihilation cross section and still take the same form (\ref{rate4}). This can also be examined in terms of thermally averaged cross
section~\cite{zhang}.

The local temperature $T(x)$ and fluid velocity $u_{\mu}(x)$ appeared in the gluon, light quark and heavy quark distribution functions are controlled by the ideal hydrodynamics~\cite{kolb},
\begin{equation}
\label{hydro}
\partial_\mu T^{\mu\nu}=0,
\end{equation}
where $T_{\mu\nu}$
is the energy-momentum tensor of the system, and we have neglected the baryon current at LHC and FCC energies. The initial time for the fluid evolution created in Pb+Pb collisions is chosen as $\tau_0=0.6$ fm/c at colliding energy $\sqrt{s_{NN}}=2.76$ and $5.5$ TeV and $0.3$ fm/c at $39$ TeV.

The equation of state for the medium is needed to close the above hydrodynamical equations.
In this work we use the Lattice QCD parametrization "s95p-v1"~\cite{pasi}, which matches the recent
Lattice QCD~\cite{hotqcd} simulation by HotQCD collaboration at high temperature (above $T_c$) to the Hadron Resnance Gas (HRG) at low temperature
with a smooth crossover transition at temperature $T_c=155$ MeV.

The final charged multiplicity is used to determine the initial entropy density of the fluid. For Pb+Pb collisions at $\sqrt{s_{NN}}=2.76$ TeV we have
$dN_{ch}/d\eta=1600$ at mid rapidity from the ALICE collaboration~\cite{alice_charge}. At higher colliding energy, we parameterize
$dN_{ch}/d\eta$ as
\begin{equation}
\frac{dN_{ch}}{d\eta}=-232.7-189.6\ln\sqrt{s_{NN}}+598.2(\sqrt{s_{NN}})^{0.217}
\label{charged}
\end{equation}
which leads to $dN_{ch}/d\eta=2000$ at $\sqrt{s_{NN}}=5.5$ TeV and 3700 at $39$ TeV. Assuming that the entropy is directly related to the final charged multiplicity,
its initial configuration is usually estimated through the two-component model~\cite{two1,two2}. We
take the mixing ratio $\alpha$ between the number density of participants $n_{part}$ and the number density of binary collisions $n_{coll}$ as
0.14, 0.16 and 0.2 for Pb+Pb collisions at $\sqrt{s_{NN}}=2.76$, $5.5$ and $39$ TeV. Being different from usual treatment, we consider here the shadowing effect on the contribution from the  hard processes,
\begin{eqnarray}
n_{coll}({\bf x}_T|{\bf b}) &=& \sigma_{NN}T_{A}({\bf x}_T)\left(1+\frac{Z_A}{T_{AB}({\bf 0})}({\cal R}_A-1)T_{A}({\bf x}_T)\right)\\
&&\times T_B({\bf x}_T-{\bf b})\left(1+\frac{Z_B}{T_{AB}({\bf 0})}({\cal R}_B-1)T_B({\bf x}_T-{\bf b})\right),\nonumber
\label{shadowingnc}
\end{eqnarray}
where the space averaged shadowing factors ${\cal R}_{A,B}$ are estimated from the EKS98 model~\cite{eks98}. The p+p inelastic cross section $\sigma_{NN}$
is taken as 62, 72 and 100 mb at $\sqrt{s_{NN}}=2.76, 5.5$ and $39$ TeV.

For the freeze out of the fluid,
we assume that the medium maintains chemical and thermal equilibrium until the energy density of the system dropping down to $60$ MeV/fm$^3$ when the interaction among hadrons ceases and their momentum distributions are fixed. By solving the hydrodynamic equations (\ref{hydro}) for the local temperature and medium velocity, and then substituting them into the rate equation (\ref{rate3}), we can numerically obtain the time evolution of the charm quark yield in the medium.

%%%%%%%%%%%%%%%%%%%%%%%%%%%%%%%%%%%%%%%%%%%%%%%%%%%%%%%%%%%%%%%%%%%%%%%
\begin{figure}[ht]
\centering
\includegraphics[width=0.4\textwidth]{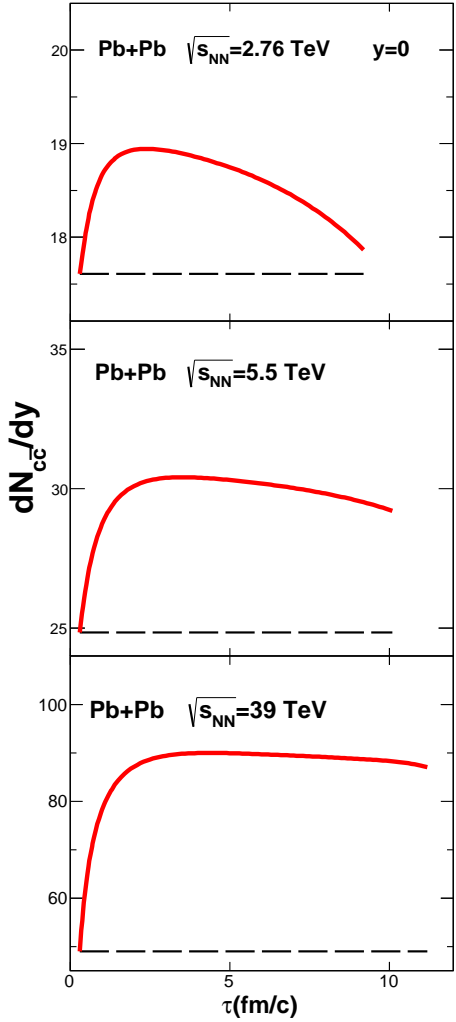}
\caption{(Color online) The time evolution of charm quark number at mid rapidity in central Pb+Pb collisions (b=0) with colliding energy $\sqrt{s_{NN}}=2.76,\ 5.5$
and $39$ TeV (from top to bottom). The shadowing effect is included, and the solid and dashed lines are the calculations with and without thermal production. }
\label{fig1}
\end{figure}
%%%%%%%%%%%%%%%%%%%%%%%%%%%%%%%%%%%%%%%%%%%%%%%%%%%%%%%%%%%%%%%%%%%%%%%
Fig.\ref{fig1} shows the charm quark number as a function of the evolution time of the medium in Pb+Pb collisions at colliding energy $\sqrt{s_{NN}}=2.76,\ 5.5$
and $39$ TeV. Since the initial gluon momentum fraction $x_g\sim 2m_T/\sqrt{s_{NN}}$ locates in the strong shadowing region~\cite{eks98}, the initially produced charm quark number (dashed lines) is significantly reduced by about $20$ - $35\%$ for $\sqrt{s_{NN}}=2.76$ - $39$ TeV. Through the thermal production inside QGP, the total charm quark number (solid lines) increases in the early period of the QGP phase and then decreases due to the charm and anti-charm quark annihilation in the later stage.
At $\sqrt{s_{NN}}=2.76$ TeV, the thermal production is not so strong and almost canceled by the inverse annihilation at the end of the QGP phase. Therefore, the consideration
of thermal charm production in medium is negligible. However, at $\sqrt{s_{NN}}=5.5$ TeV, the QGP medium becomes hotter and survives longer, and the thermal
charm production becomes important and leads to a visible increase of the total charm quark number. At the end, the thermal production induced increase is still $15\%$. This sizeable enhancement agrees with the previous calculations like \cite{zhang} using a fire-cylinder model for the medium evolution. At the FCC energy $\sqrt{s_{NN}}=39$ TeV, the initial temperature of the medium in
central collisions is $T_0$= 840 MeV, and the densely populated gluons at so high temperature in QGP makes the thermal charm production more efficient. In this case, the total charm quark number increases exponentially in the very beginning of the QGP medium and keeps as almost a constant in the later evolution. At the end of the QGP, the enhancement reaches $50\%$.

Due to the heavy mass, charmonium evolution in QGP can be controlled by classical transport approaches. At mid rapidity in heavy ion collisions, the Boltzmann equation
can be used to describe the charmonium distribution function $f_{\Psi}({\bf x}_T,{\bf p}_T,\tau|{\bf b})$ in transverse phase space $({\bf x}_T,{\bf p}_T)$ at time $\tau$ and fixed impact
parameter ${\bf b}$,
\begin{equation}
\partial_{\tau}f_{\Psi}+{\bf v}_{\Psi}\cdot \nabla f_{\Psi}=-\alpha_{\Psi}f_{\Psi}+\beta_{\Psi},
\label{transport}
\end{equation}
where $\Psi$ stands for $J/\psi$, $\chi_c$ and $\psi'$ to include the feed-down~\cite{feeddown} contribution from excited states $\chi_c$ and $\psi'$ to ground state $J/\psi$, ${\bf v}_{\Psi}={\bf p}_T/\sqrt{p^2_T+m^2_{\Psi}}$
is the $\Psi$ transverse velocity. On the right hand side, the loss and gain terms $\alpha({\bf x}_T,{\bf p}_T,\tau|{\bf b})$ and $\beta({\bf x}_T,{\bf p}_T,\tau|{\bf b})$
represent charmonium dissociation and regeneration in the hot medium. The elastic scattering is neglected here, since the $\Psi$ masses are much larger than the typical medium
temperature. Considering the gluon dissociation $g+\Psi \to c+\bar c$ as the main dissociation process, the loss term is given by~\cite{zhu,yan}
\begin{equation}
\label{loss}
\alpha_{\Psi}=\frac{1}{2E_T}\int{d^3{\bf k}\over(2\pi)^3 2E_g}\sigma_{g\Psi}({\bf p},{\bf k},T)4F_{g\Psi}({\bf p},{\bf k})f_g({\bf k},T,u_\mu),
\end{equation}
where $E_g$ is the gluon energy and $F_{g\Psi}=\sqrt{(p k)^2-m_\Psi^2m_g^2}=p k$ the flux factor. For the dissociation cross section
$\sigma_{g\Psi}({\bf p},{\bf k},0)$ in vacuum at $T=0$, one can use the result from the operator production expansion (OPE) method with a perturbative Coulomb interaction~\cite{ope1,ope2,ope3,ope4}.
When going to high temperature medium where the charmonium states become loosely bound, the OPE method is no longer valid. We estimate the temperature effect by taking into account the geometrical relation between the cross sections in medium and in vacuum~\cite{liu,tang},
\begin{equation}
\label{crosssection}
\sigma_{g\Psi}({\bf p},{\bf k},T)={\langle r^2\rangle_\Psi(T)\over \langle r^2\rangle_\Psi(0)}\sigma_{g\Psi}({\bf p},{\bf k},0),
\end{equation}
where $\langle r^2\rangle_\Psi(T)$ is the averaged charmonium radius square which can be calculated by the potential model~\cite{pot1} with lattice simulated heavy
quark potential~\cite{petreczky} at finite temperature. The divergence of $\langle r^2\rangle_\Psi(T)$ defines the charmonium dissociation temperature $T_d$
above which the charmonium state $\Psi$ will melt due to the color screening.

Using detailed balance between the dissociation and regeneration processes~\cite{thews}, we can get the regeneration cross section.
To obtain the regeneration rate $\beta$, we need the charm quark phase space distribution $f_{c(\bar c)}$ which includes the contribution from the thermal charm production described above.

The initial condition for the charmonium transport equation is obtained~\cite{zhou} by taking the geometric superposition of free p+p collisions along with the modifications from the cold nuclear matter effects like Cronin effect~\cite{cronin} and nuclear shadowing effect~\cite{shad}.

%%%%%%%%%%%%%%%%%%%%%%%%%%%%%%%%%%%%%%%%%%%%%%%%%%%%%%%%%%%%%%%%%%%%%%%
\begin{figure}[ht]
\centering
\includegraphics[width=0.4\textwidth]{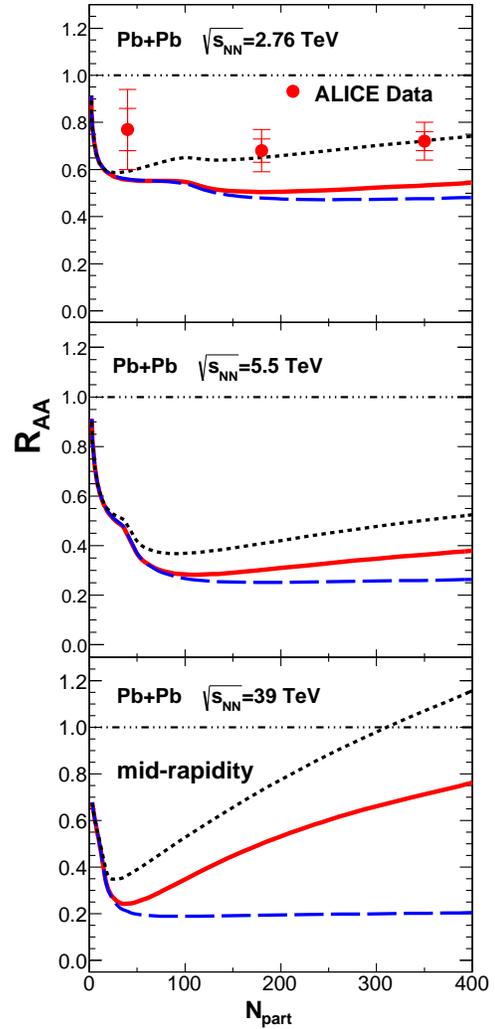}
\caption{(Color online) The centrality dependence of $J/\psi$ nuclear modification factor $R_{AA}(N_{part})$ at mid rapidity in Pb+Pb collisions with colliding energy $\sqrt{s_{NN}}=2.76,\ 5.5$ and $39$ TeV (from top to bottom). The solid and dashed lines are the calculations with and without thermal production. To see the contribution from the shadowing effect, we show also the calculation with thermal production but without shadowing effect (dotted lines). The LHC data are from the ALICE Collaboration~\cite{alice2}. }
\label{fig2}
\end{figure}
%%%%%%%%%%%%%%%%%%%%%%%%%%%%%%%%%%%%%%%%%%%%%%%%%%%%%%%%%%%%%%%%%%%%%%%
Fig.\ref{fig2} shows the transverse momentum integrated ($0<p_t<30$ GeV) $J/\psi$ nuclear modification factor $R_{AA}$ as a function of the number of participant nucleons $N_{part}$ at mid rapidity in Pb+Pb collisions with colliding energy $\sqrt{s_{NN}}=2.76,\ 5.5$ and $39$ TeV. The thermal production of charm quarks does not change the initial charmonium production before the medium formation but enhances the charmonium regeneration inside the hot medium. At the current LHC energy $\sqrt{s_{NN}}=2.76$ TeV, the weak thermal charm production slightly enhances the charmonium regeneration and in turn leads to a small $J/\psi$ enhancement even in very central collisions, see the difference between the calculations with (solid line) and without (dashed line) thermal production shown in the upper panel of Fig.\ref{fig2}. Since the degree of the nuclear shadowing is still an open question and there is not yet precise calculation of the shadowing factor, we show, as a comparison, also the calculation with thermal production but without shadowing in Fig.\ref{fig2} (dotted lines). Since the shadowing effect suppresses both the initial production and regeneration, the dotted line is always above the solid line. From the comparison with the ALICE data~\cite{alice2}, the calculation without shadowing effect looks more close to the data.

At $\sqrt{s_{NN}}=5.5$ TeV, there exists a wide plateau of $R_{AA}$ in semi-central and central collisions, when the thermal charm production is excluded, similar to the structure at RHIC energy~\cite{liu}. The thermal charm production which increases with centrality breaks the plateau structure, and the charmonium yield goes up sizeably with the number of participants. Considering the fact that the charmonium regeneration is proportional to the charm quark number square, a small change in the charm quark number by thermal production can lead to a remarkable charmonium enhancement. For instance, in very central collisions at $\sqrt{s_{NN}}=5.5$ TeV, the charm quark enhancement is about $15\%$, but the charmonium enhancement becomes $(0.37-0.26)/0.26 \sim 40\%$.

Very different from the case at colliding energies $\sqrt{s_{NN}}=2.76$ and 5.5 TeV, the thermal charm production plays a dominant role in charmonium production at FCC energy $\sqrt{s_{NN}}$=39 TeV. Since the fireball created at FCC is extremely hot, the size is so large, and the life time is so long, the initially produced charmonia are almost all eaten up by the hot medium, and the charmonium production is controlled by the regeneration, and therefore the contribution from the thermal charm production to the charmonium yield is largely amplified, see the bottom panel of Fig.\ref{fig2}. In very central collisions, the thermal charm production makes the $J/\psi$ $R_{AA}$ increase from $0.2$ to $0.75$, being enhanced by a factor of 3! When the shadowing is switched off, the $R_{AA}$ can even be larger than one. The other significant signal of the thermal charm production is the deep valley of $R_{AA}$ located at $N_{pant}\sim 30$, due to the competition between the initial production in peripheral collisions and strong thermal charm production in semi-central and central collisions.

%%%%%%%%%%%%%%%%%%%%%%%%%%%%%%%%%%%%%%%%%%%%%%%%%%%%%%%%%%%%%%%%%%%%%%%
\begin{figure}[ht]
\centering
\includegraphics[width=0.4\textwidth]{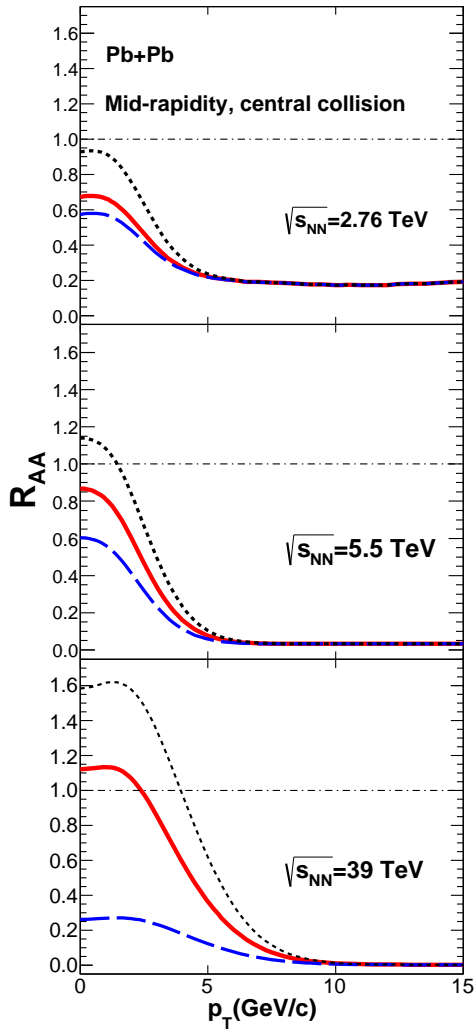}
\caption{(Color online) The differential $J/\psi$ $R_{AA}(p_t)$ at mid rapidity in central Pb+Pb collisions (b=0) with colliding energy $\sqrt{s_{NN}}=2.76,\ 5.5$ and $39$ TeV (from top to bottom). The solid and dashed lines are the calculations with and without thermal production. To see the contribution from the shadowing effect, we show also the calculation with thermal production but without shadowing effect (dotted lines). }
\label{fig3}
\end{figure}
%%%%%%%%%%%%%%%%%%%%%%%%%%%%%%%%%%%%%%%%%%%%%%%%%%%%%%%%%%%%%%%%%%%%%%%
The charmonium transverse momentum distribution is more sensitive to the production mechanism and the medium properties. The two charmonium production mechanisms, namely the initial production and the later regeneration, play roles in different $p_t$ regions. For the initially produced charmonia, the low $p_t$ part is almost all absorbed by the hot medium, while the high $p_t$ part can survive the medium due to the leakage effect~\cite{leakage,zhou2}. Since the regeneration process happens in the later stage of the medium evolution, the regenerated charmonia carry low energy and mainly distribute in the low $p_t$ region. Therefore, the thermal charm production which contributes only to the regeneration will enhance the charmonium yield in low $p_t$ region. The $p_t$ dependence of the $J/\psi$ $R_{AA}$ at mid rapidity is shown in Fig.\ref{fig3} for central Pb+Pb collisions (b=0) at different colliding energy. At the current LHC energy $\sqrt{s_{NN}}=2.76$ TeV, the high $p_t$ region ($p_t>5$ GeV/c) is dominated by the initial production, and the regeneration sourced from those initially produced charm quarks controls the low $p_t$ region. The thermal charm production leads to a remarkable enhancement at very low $p_t$, see the comparison between the dashed and solid lines in the top panel of Fig.\ref{fig3}. Without considering the shadowing effect, the thermal production induced enhancement becomes more remarkable.

With increasing colliding energy, the initially produced charmonia are more suppressed by the hotter medium, which results in a smaller $R_{AA}$ at high $p_t$. On the other hand, only those regenerated charmonia in the temperature region $T_c<T<T_d$ can survive in the QGP phase, those regenerated charmonia above the dissociation temperature $T_d$ will be immediately dissociated by the hot medium. This is the reason why the $R_{AA}$ at low $p_t$ is very small ($\sim 0.25$) at FCC energy when the thermal charm production is excluded, see dashed lines in Fig.\ref{fig3}. However, when the thermal charm production is turned on, the $J/\psi$ yield at low $p_t$ goes up monotonously with increasing colliding energy. Especially, the $J/\psi$ suppression ($R_{AA}<1$) at $\sqrt{s_{NN}}=2.76$ and 5.5 TeV becomes enhancement ($R_{AA}>1$) at the FCC energy. When the shadowing is switched off, the $R_{AA}$ can even reach 1.6, the enhancement by the thermal charm production is tremendous, see the dotted line in the bottom panel.

In summary, the effect of thermal charm production in QGP phase on the charmonium production in high energy nuclear collisions at LHC and FCC energies is investigated. There might also be minor contributions from pre-thermal equilibrium stage for thermal charm production, here in this exploratory work we neglect it assuming very fast thermalization for the whole system. We calculated the global and differential nuclear modification factors $R_{AA}(N_{part})$ and $R_{AA}(p_t)$ for $J/\psi$s, by solving the kinetic equations for charm quarks and charmonia in the QGP phase.
While the thermal charm production is still weak at the current LHC energy $\sqrt{s_{NN}}=2.76$ TeV, it becomes sizeable at $\sqrt{s_{NN}}=5.5$ TeV and significant at the FCC energy. As a consequence, the source for charmonium production changes from initially created charm quarks at SPS, RHIC and LHC to thermally produced charm quarks at FCC, and the charmonium production in Pb+Pb collisions at FCC is completely controlled by the thermal charm production. This is manifested in the following three aspects: 1) The $J/\psi$ yield in central collisions is enhanced by a factor of $4$, 2) There appears a deep valley of global $R_{AA}$ located at peripheral collisions, and 3) the $J/\psi$ suppression ($R_{AA}<1$) at low transverse momentum at SPS, RHIC and LHC energies becomes $J/\psi$ enhancement ($R_{AA}>1$) at FCC energy.

\noindent {\bf Acknowledgement:}  Kai Zhou thanks the Frankfurt international graduate school for science for its financial support. The work is supported by the NSFC and MOST grant Nos. 11335005, 11575093, 2013CB922000 and 2014CB845400.


\begin{thebibliography}{99}
\bibitem{satz} T.~Matsui and H.~Satz, Phys. Lett. {\bf B178}, 417(1986).
\bibitem{review} A.~Andronic {\it et al.}, arXiv:1506.03981.
\bibitem{review_zhuang} Y.~Liu, K.~Zhou and P.~Zhuang, Int. J. Mod. Phys. {\bf E24} 11, 1530015(2015).
\bibitem{hufner} J.~Huefner and B.~Kopeliovich, Phys. Lett. {\bf B445},223(1998).
\bibitem{vogt} S.~Gavin and R.~Vogt, Phys. Rev. Lett. {\bf 78}, 1006(1997).
\bibitem{zhu} X.~Zhu, P.~Zhuang and N.~Xu, Phys. Lett. {\bf B607}, 107(2005).
\bibitem{pbm} P.~Braun-Munzinger and J.~Stachel, Phys. Lett. {\bf B490}, 196(2000).
\bibitem{thews} R.~Thews, M.~Schroedter and J.~Rafelski, Phys. Rev. {\bf C63}, 054905(2001).
\bibitem{rapp} L.~Grandchamp, R.~Rapp and G.~Brown, Phys. Rev. Lett. {\bf 92},212301(2004).
\bibitem{yan} L.~Yan, P.~Zhuang and N.~Xu, Phys. Rev. Lett. {\bf 97}, 232301(2006).
\bibitem{taesoo} T.~Song, W.~Park, S.~Lee, Phys. Rev. {\bf C81}, 034914(2010).
\bibitem{strikland} M.~Strickland, Phys. Rev. Lett. {\bf 107}, 132301(2011).
\bibitem{blaizot} J.~Blaizot, D.~Boni, P.~Faccioli, G.~Garberoglio, Nucl. Phys. {\bf A946} 49-88(2016).
\bibitem{katz} R.Katz, P.Gossiaux, arXiv:1504.08087.
\bibitem{fcc} N.~Armesto, A.~Dainese, D.~d'Enterria, S.~Masciocchi, C.~Roland, C.~Salgado, M.~van Leeuwen and U.~Wiedemann, Nucl. Phys. {\bf A931}, 1163(2014).
\bibitem{lo1} P.~Levai, B.~Muller and X.~Wang, Phys. Rev. {\bf C51}, 3326(1995).
\bibitem{lo2} B.~Kaempfer and O.~Pavlenko, Phys. Lett. {\bf B391}, 185(1997).
\bibitem{lo3} J.~Uphoff, O.~Fochler, Z.~Xu and C.~Greiner, Phys. Rev. {\bf C82}, 044906(2010).
\bibitem{zhang} B.~Zhang, C.~Co and W.~Liu, Phy. Rev. {\bf C77},  024901(2008).
\bibitem{quench1} B.~Abelev, {\it et al.}, [STAR Collaboration], Phys. Rev. Lett. {\bf 98}, 192301(2007).
\bibitem{quench2} A.~Dainese, [ALICE Collaboration], Pos. ICHEP2012, 417(2012).
\bibitem{dv1} A.~Adare, {\it et al.}, [PHENIX Collaboration], Phys. Rev. Lett. {\bf 98}, 172301(2007).
\bibitem{dv2} Z.~Conesa del Valle, [ALICE Collaboration], Nucl. Phys. {\bf A904-905}, 178c(2013).
\bibitem{taesoo2} T.~Song, W.~Park and S.~Lee, Phys. Rev. {\bf C84}, 054903(2011).
\bibitem{bjorken} J.~Bjorken, Phys. Rev. {\bf D27}, 140(1983).
\bibitem{fonll} M.~Cacciari, M.~Greco and P.~Nason, JHEP {\bf 9805}, 007(1998); M.~Cacciari, S.~Frixione and P.~Nason, JHEP {\bf 0103}, 006(2001).
\bibitem{vogt2} R.~Vogt, Phys. Rev. {\bf C71}, 054902(2005).
\bibitem{eks98} K.~Eskola, V.~Kolhinen and C.~Salgado, Eur. Phys. J. {\bf C9}, 61(1999).
\bibitem{nason} P.~Nason, S.~Dawson, and R.~Ellis, Nucl. Phys. B {\bf 303},607 (1988); Nucl. Phys. {\bf B327}, 49(1989).
\bibitem{mnr}  M.L.~Mangano, P.~Nason and G.~Ridolfi, Nucl. Phys. {\bf B373}, 295(1992).
\bibitem{mass} E.~Braaten and R.~Pisarski, Phys. Rev. {\bf D45}, 1827(1992).
\bibitem{plumari} S.~Plumari, W.M.~Alberico, V.~Greco and C.Ratti, Phys. Rev. {\bf D84}, 094004(2011).
\bibitem{kolb} P.~Kolb, J.~Sollfrank and U.~Heinz, Phys. Rev. {\bf C62}, 054909(2000).
\bibitem{pasi} P.~Huovinen and P.~Petreczky, Nucl. Phys. {\bf A837}, 26(2010).
\bibitem{hotqcd} A.~Bazavov, {\it et al.} [HotQCD Collaboration], Phys. Rev. {\bf D80}, 014504(2009).
\bibitem{alice_charge} K.~Gulbrandsen [ALICE Collaboration], J. Phys. Conf. Ser.{\bf 446}, 012027(2013).
\bibitem{two1} D.~Kharzeev and M.~Nardi, Phys. Lett. {\bf B507}, 121(2001).
\bibitem{two2} P.~Bozek,  M.~Chojnacki, W.~Florkowski and B.~Tomasik, Phys. Lett. {\bf B694}, 238(2010).
\bibitem{feeddown} A.~Zoccoli, {\it et al.} [HERA-B Collaboration], Eur. Phys. J. {\bf C43}, 179(2005).
\bibitem{ope1} G.~Bhanot and M.~Peskin, Nucl. Phys. {\bf B156}, 365(1979); Nucl. Phys. {\bf B156}, 391(1979).
\bibitem{ope2} F.~Arleo, {\it et al.}, Phys. Rev. {\bf D65}, 014005(2002).
\bibitem{ope3} YS.~Oh, S.~Kim and S.~Lee, Phys. Rev. {\bf C65}, 067901(2002).
\bibitem{ope4} X.~Wang, Phys. Lett. {\bf B540}, 62(2002).
\bibitem{liu} Y.~Liu, Z.~Qu, N.~Xu and P.~Zhuang, Phys. Lett. {\bf B678}, 72(2009).
\bibitem{tang} Z.~Tang, K.~Zhou, N.~Xu and P.~Zhuang, J. Phys. {\bf G41}, 124006(2014).
\bibitem{pot1} H.~Sata, J. Phys. {\bf G32}, R25(2006).
\bibitem{petreczky} P.~Petreczky, J. Phys. {\bf G37}, 094009(2010).
\bibitem{zhou} K.~Zhou, N.~Xu, Z.~Xu and P.~Zhuang, Phys. Rev. {\bf C89}, 054911(2014).
\bibitem{cronin} J.~Cronin, {\it et al.}, Phys. Rev. {\bf D11}, 3105(1975); J.~Hufner, Y.~Kurihara and H.~Pirner, Phys. Lett. {\bf B215}, 218(1988).
\bibitem{shad} A.~Mueller and J.~Qiu, Nucl. Phys. {\bf B268}, 427(1986).
\bibitem{alice2} Pereira Da Costa Hugo {\it et al}. (ALICE Collaboration), arXiv:1110.1035, AIP Conf. Proc. 1441, 859 (2012).
\bibitem{leakage} J.~Hufner and P.~Zhuang, Phys. Lett. {\bf B559}, 193(2003).
\bibitem{zhou2} K.~Zhou, N.~Xu and P.~Zhuang, Nucl. Phys. {\bf A834}, 249c(2009).
\end{thebibliography}
\end{document}